\def\versionno{ v4.6 }

\catcode`\@=11
\newif\if@fewtab\@fewtabtrue
{\count255=\time\divide\count255 by 60
\xdef\hourmin{\number\count255} 
\multiply\count255 by-60\advance\count255 by\time
\xdef\hourmin{\hourmin:\ifnum\count255<10 0\fi\the\count255}}
\def\ps@draft{\let\@mkboth\@gobbletwo
    \def\@oddfoot{\hbox to 7 cm{\tiny \versionno
       \hfil}\hskip -7cm\hfil\rm\thepage \hfil {\tiny\draftdate}}
    \def\@oddhead{}
    \def\@evenhead{}\let\@evenfoot\@oddfoot}
\def\draftdate{\number\month/\number\day/\number\year\ \ \ \hourmin }

\global\def\draftcontrol{0}
\def\citen#1{\if@filesw \immediate\write \@auxout {\string\citation{#1}}\fi%
\@tempcntb\m@ne \let\@h@ld\relax \def\@citea{}%
\@for \@citeb:=#1\do {\@ifundefined {b@\@citeb}%
    {\@h@ld\@citea\@tempcntb\m@ne{\bf ?}%
    \@warning {Citation `\@citeb ' on page \thepage \space undefined}}%
    {\@tempcnta\@tempcntb \advance\@tempcnta\@ne
    \setbox\z@\hbox\bgroup\ifcat0\csname b@\@citeb \endcsname \relax
    \egroup \@tempcntb\number\csname b@\@citeb \endcsname \relax
    \else \egroup \@tempcntb\m@ne \fi \ifnum\@tempcnta=\@tempcntb
    \ifx\@h@ld\relax \edef \@h@ld{\@citea\csname b@\@citeb\endcsname}%
    \else \edef\@h@ld{\hbox{--}\penalty\@highpenalty
    \csname b@\@citeb\endcsname}\fi
    \else \@h@ld\@citea\csname b@\@citeb \endcsname \let\@h@ld\relax \fi}%
\def\@citea{,\penalty\@highpenalty\hskip.13em plus.13em minus.13em}}\@h@ld}
\def\@citex[#1]#2{\@cite{\citen{#2}}{#1}}%
\def\@cite#1#2{\leavevmode\unskip\ifnum\lastpenalty=\z@\penalty\@highpenalty\fi%
  \ [{\multiply\@highpenalty 3 #1%
  \if@tempswa,\penalty\@highpenalty\ #2\fi}]}   %
\makeatother 
\catcode`\@=12

\def\be            {\begin{eqnarray}}
\def\bearl         {\begin{array}{l}}
\def\bearll        {\begin{array}{ll}}

\def\chii          {\raisebox{.15em}{$\chi$}}

\def\complex       {\mbox{$\mathbb C$}}

\global\def\draftcontrol{0}

\def\eE            {{\rm e}}
\def\ee            {\end{eqnarray}}
\def\eear          {\end{array}}

\newcommand\erf[1] {(\ref{#1})}

\def\ii            {{\rm i}}

\newcommand\labl[1]{\label{#1}\ee \ifnum\draftcontrol=1
                   \mbox{ }\\[-12 mm]\query{#1}\\[5 mm] \fi}
\newcommand\Labl[2]{\label{#1#2}\ee \ifnum\draftcontrol=1
                   \mbox{ }\\[-12 mm]\query{#1#2}\\[5 mm] \fi}

\long\def\query#1{\hskip 0pt{\vadjust{\everypar={}\small\vtop to 0pt{\hbox{}%
     \vskip -13pt\rlap{\hbox to 49.0pc{\hfil{\vtop{\hsize=8pc\tolerance=6000%
     \hfuzz=.5pc\rightskip=0pt plus 3em\noindent#1}}}}\vss}}}}%

\def\ints         {\mbox{$\mathbb Z$}}

\newcommand\sect[1]{\section{#1}\setcounter{equation}{0}}

\def\su            {\mbox{$\mathfrak{su}$}}

\newcommand\tfrac[2]  {\mbox{\large$\frac{#1}{#2}$}}

\documentclass[12pt]{article}

\usepackage{amssymb,amsfonts,epsf,color,colordvi,fancybox}
\usepackage[dvips]{graphics} 
\begin{document}

\begin{flushright}  {~} \\[-1cm] {\sf LPTHE-P03-05} \\[1mm]
{\sf March 2003} \end{flushright}

  \begin{center} \vskip 22mm
  {\Large\bf D-BRANES IN LENS SPACES}
 \\[22mm]
 {\large Pedro Bordalo }\,$^1$ \ and \ {\large Albrecht Wurtz }\,$^2$
 \\[13mm]
 $^1\;$ LPTHE, Universit\'e Paris VI~~~~{}\\
 4 place Jussieu\\
 F\,--\,75 252 Paris Cedex 05
 \\[7mm]
 $^2\;$ Institutionen f\"or fysik~~~~{}\\
    Universitetsgatan 5\\ S\,--\,651\,88\, Karlstad\\[5mm]
  \end{center} \vskip 23mm

\begin{quote}{\bf Abstract}
\\[3mm]
We realize the CFT with target a lens space $SU(2)/\ints_{l}$ as a simple current construction. This 
allows us to compute the boundary states and the annuli coefficients, and in particular 
to study the B-type branes, in purely algebraic terms. 
Several issues, like the appearance of fractional branes and symmetry breaking boundary 
conditions, can be addressed more directly
in this approach than in a more geometric treatment. 

\end{quote}

\newpage

\sect{Introduction}

Boundary conditions in rational conformal field theories have been the focus of intense study
over the past few years.  For a very large class of theories, those with partition function of 
{\it simple current type}, the boundary conditions are known explicitly \cite{FHSSW}.
The best-known examples of these are WZW models on non-simply connected groups, 
where there is a target space interpretation of both the bulk model and of its boundary 
conditions.

Recently, there has been interest in theories whose target spaces are  quotients 
of Lie groups by discrete subgroups \cite{MMS1,MMS2,QuSch}.
The simplest non-trivial example of such theories are those with the target being a 
lens space $L_{k_1} = SU(2) / \mathbb{Z}_{k_1}$. Geometrically,
this space is obtained by quotienting the group manifold $SU(2)$ by the left action of the 
subgroup $\mathbb{Z}_{k_1}$ of one of its maximal tori, so 
that the elements $g$ and $\eE ^{\frac{2\pi \ii }{k_1}H }g$ of $SU(2)$
are identified, for $k_1$ integer and $H$ the generator of a maximal torus.   
This $\mathbb{Z}_{k_1}$-action has no fixed points, so the lens spaces are smooth 
manifolds; they inherit a metric and volume form from the
translation invariant metric and volume form of the covering space $SU(2)$.
The quotienting is therefore easily implemented in the sigma-model description of any $SU(2)_k$
WZW model.  However, in order that integrality of the Wess-Zumino term 
is preserved, and since $\mathbb{Z}_{k_1}$ is acting freely, the level $k$ of the
$SU(2)$ WZW model must be a multiple of the order of $\mathbb{Z}_{k_1}$ \cite{GPS}, 
i.e. $k=k_1 k_2$. We denote the  resulting lens space CFT by $\mathcal{L}_{k,k_1}$.  
Performing a T-duality transformation along the maximal torus that has $\ints_{k_1}$ 
as a subgroup on the CFT $\mathcal{L}_{k,k_1}$ gives the theory $\mathcal{L}_{k,k_2}$ \cite{MMS1}. Lens spaces have for example been used as backgrounds for string propagation,
and in particular they appeared in the bosonization of the near horizon limit 
of a four-dimensional black hole \cite{GPS}.

Boundary conditions in lens spaces 
were studied in \cite{MMS1} from a geometric perspective, different from ours.   
It turns out that the CFT can be regarded as
an asymmetric orbifold with phenomena such as fractional branes, as was known 
\cite{FFFS} for the special case $SO(3) = SU(2)/\mathbb{Z}_2$ 
(in which the orbifold is not asymmetric) which is described by the CFT $\mathcal{L}_{k,2}$.  
Furthermore, via T-duality one obtains a different kind of branes, called B-branes \cite{MMS1}. 

\smallskip

In this note we show that the CFT $\mathcal{L}_{k,k_1}$ can be described as a simple current 
construction from the CFT $PF \otimes U(1)$ at level $k$, 
with non-trivial discrete torsion. Here $PF$ denotes the parafermions  
and $U(1)$ is a rational free boson.
The formalism is in particular adapted to describing the fractional branes, 
thus allowing us to recover and generalize some results of \cite{MMS1} regarding 
boundary states for A- and B branes and annulus amplitudes, and to resolve 
some issues which in the CFT description involves fixed points of 
the simple current action, that are hard to understand with purely geometric methods. 
Our approach yields all boundary conditions that preserve the $PF \otimes U(1)$ chiral 
algebra on the same footing, i.e.\ in the case of $SU(2)$ both the $SU(2)$-symmetric
boundary conditions that were studied in \cite{MMS1} and the $SU(2)$-breaking ones 
that in the large $k$ limit correspond \cite{FFFS} to certain twisted conjugacy classes.

One can think of two natural extensions of the problems approached in this paper;
both directions represent an interesting challenge.
The first is to construct more general
boundary states in $SU(2)$ using more general $U(1)$ boundary states;
this requires a better understanding of $U(1)$ boundary states, which is a problem
of independent interest. 
The second direction is to study theories obtained by
modding out non-abelian subgroups of $SU(2)$, including their fixed point structures.  
The inverse operation to forming such a non-abelian orbifold is an extension that
generalizes simple current extensions; methods for performing such constructions
explicitly have been introduced only recently \cite{FRS}. In the case at hand,
one should in particular perform such constructions starting from
the theory $PF \otimes U(1)/\ints_2$ instead of
$PF \otimes U(1)$, with $U(1)/\ints_2$ denoting the $\ints_2$ orbifold of the free boson.
This would in particular 
allow one to
describe the $A$- and $B$-type boundary conditions on the same footing.


\sect{Lens spaces as CFTs of simple current type}

\paragraph{Simple current constructions}

To study branes in lens spaces, we start by describing the closed string 
spectrum of the CFTs $\mathcal{L}_{k,k_1}$, 
that is, their torus partition functions.
The simple current modular invariants that give rise to consistent
conformal field theories have been classified ($\!\!$\cite{KS}, see also \cite{GRS})
and indeed
constitute the vast majority of known rational CFTs. 
Starting from a chiral RCFT with chiral algebra $\mathcal{A}$ and irreducible 
$\mathcal{A}$-representations labelled by $\lambda$, with characters $\chii_\lambda$,
a modular invariant is of simple current type if all its terms are of the form  
$\chii_{\lambda}\bar{\chii}_{J*\lambda^{\dagger}}$ with $J$ a simple current and $*$ the fusion product. These theories 
are characterized (up to exceptional simple current invariants, see for instance \cite{FSS3}) by the choice of a simple current group $\mathcal{G}$ 
and by a certain matrix $X$. 
Here $\mathcal{G}$ is a finite abelian group (with respect to the 
fusion product); it can be written as
$\mathcal{G} \cong \mathbb{Z}_{n_1}\times \cdots \times \mathbb{Z}_{n_q}$ with 
$n_{i+1}|n_i$.
Picking a set of generators $J_i$ of $\mathcal{G}$, one defines the off-diagonal part of a symmetric 
$q\times q$ matrix $R$ by the relative monodromy charges:
$R_{ij} := Q_{J_i} (J_j) := \Delta(J_i) + \Delta(J_j) - \Delta(J_i*J_j)$ mod $1$. 
The diagonal part of $R$ is required to satisfy $2\Delta(J_k)=(n_k{-}1)R_{kk}$ mod $2$. 
The symmetric part of the $q\times q$ matrix $X$ is then fixed by $X+X^t = R \bmod 1$,
while its antisymmetric part, called discrete torsion, is constrained by ${\rm gcd}(n_i,n_j)X_{ij}\in\ints$.  
(Note that this 
is not necessarily the same concept as the discrete torsion that arises in 
geometric orbifolds \cite{vafa}.)

We denote the theory constructed from $\mathcal{A}$ with simple current 
group $\mathcal{G}$ and matrix $X$ by $\{\mathcal{A}*_X \mathcal{G} \}$.
Denoting an element of $\mathcal{G}$ by $J^{\vec{s}}= \Pi_{i=1}^q J_i^{s_i}$, 
the modular invariant torus partition function
of the theory $\{\mathcal{A}*_X \mathcal{G} \}$ is the combination
\be
Z = \sum_\lambda \sum_{J^{\vec{s}}\in\mathcal{G}} \left(\prod_{i=1} ^q
    \delta^1 (Q_{J_i} (\lambda)+X_{ij} s_i)\right)
    \chii_\lambda\, \bar{\chii}_{J^{\vec{s}}\lambda^{\dagger}}  \label{modinv}
\ee
of $\mathcal{A}$-characters \cite{KS}. Here $\delta^1(x)=1$ if $x\in\ints$ and $0$ else.
The left and right kernels of the matrix $X$ determine the extension of the left and right
chiral algebras, respectively; thus when they are different,
the modular invariant (\ref{modinv}) is left-right asymmetric.  

\paragraph{SU(2)}

The parafermion theory $PF_k$ can be constructed as a $SU(2)/U(1)$ coset 
model at level $k$, as described e.g.\ in \cite{GQ}. 
The primary fields of $PF_k$ are labelled by pairs of integers
$(j,n)$ with $0\,{\leq}\, j \,{\leq}\, k$, $\,0\,{\leq}\, n \,{<}\,2k$,
subject to the selection rule $2|j{+}n$, i.e.\ $j$ and $n$ must have the 
same parity.
Furthermore, the labels $(j,n)$ and $(k{-}j,n{+}k)$ describe one and the
same field. This is known as field identification and will be denoted by
$(j,n)\sim (k{-}j,n{+}k)$;
the selection rule as well as the field identification arise naturally in the $SU(2)/U(1)$ coset
construction. The conformal dimension of a primary field labelled $(j,n)$ is 
$\Delta(j,n)=\frac{j(j+2)}{4(k+2)}-\frac{n^2}{4k}$ for $n\,{\le}\,k$ and 
$\Delta(j,n)=\frac{j(j+2)}{4(k+2)}-\frac{(n-2k)^2}{4k}$ for $n\,{>}\,k$.

Our conventions for the rational free boson CFT $U(1)_k$ are that the primary fields
are labelled by $\,0\,{\leq}\, m \,{<}\,2k$, in terms of which the conformal 
dimensions are $\Delta_{m}=\frac{m^2}{4k}$ for $m\,{\le}\,k$ and 
$\Delta(m)=\frac{(m-2k)^2}{4k}$ for $m\,{>}\,k$.
In the tensor product theory 
$PF_k \otimes U(1)_k$ we have a simple current group $\mathcal{G}_k \cong \mathbb{Z}_k$ that is
generated by the field $(0,2,2)$ which acts as $(0,2,2)*(j,n,m)=(j,n+2,m+2)$.
Using the fact that the discrete torsion of a cyclic group is 
necessarily trivial, and that
\be
\sum_{m\in \mathbb Z_{2k}\atop{2|m+j}}\,\chii_{(j,m)}^{PF_k}\,\,\chii_{(m)}^{U(1)_k}
=\chii^{SU(2)_k}_{j}, \label{chardec}
\ee
one can check that, when applied to the $PF \otimes U(1)$
theory with this choice of simple current group, 
the prescription (\ref{modinv}) gives us the charge conjugation $SU(2)$ partition function, 
hence $\{(PF_k\otimes U(1)_k)* \mathcal{G}_k \}=SU(2)_k$.

\paragraph{Lens spaces} 

The torus partition functions
for the lens space CFTs have been obtained \cite{GPS,MMS1} by requiring level 
matching of the twisted vertex operators on the orbifold $L_{k_1}$, as well as the right $su(2)_k$ symmetry to be preserved, as
\be
Z\left( \mathcal{L}_{k,k_1} \right) = \sum_{j=0} ^k
 \left( \sum_{n-n' = 0 \bmod 2k_2\atop{n+n' = 0 \bmod 2k_1}}
 \chii_{jn}^{PF_k}\chii_{n'}^{U(1)_k}\right)  \bar{\chii}^{SU(2)_k} _{j} \,, \label{partition}
\ee
where $n{=}j\bmod 2$ and $n,n'{=}\,0, \ldots, 2k{-}1$.
In view of \erf{chardec}, this is a combination of parafermion and $U(1)$ characters,
with the left and right combinations of labels all connected by the action of suitable
simple currents. Any such partition function is of simple current type (see \cite{KS}); thus
the lens space CFT can be obtained as a simple current construction 
from the $PF_k\times U(1)_k$ theory. Furthermore, since \erf{partition} is 
left-right asymmetric, we need at least a $2\times 2$ matrix $X$, and hence a non-cyclic 
simple current group with at least two factors, $\ints_{l_1}\times\ints_{l_2}$. 
And since we have $SU(2)_k$-characters on the right, we need in particular all those 
currents that appear in the construction of the $SU(2)_k$ theory, hence we 
let one of these factors be $\mathcal{G}_k \cong \ints_k$. 
One can also check that (\ref{partition}) with $k_1=2$, 
i.e.\ $\mathcal L_{k,2}$ gives the $SO(3)=SU(2)/\ints_2$ 
WZW theory at level $k/2$, which is a $SU(2)*\ints_2$ simple current construction.
It is therefore natural to try a simple current group 
$\mathcal{G}_{k,k_1} \cong \mathbb{Z}_{k} \times \mathbb{Z}_{k_1}$ with $k_1|k$. 
We take again the generator $J_1=(0,2,2)$
for the $\ints_k$ factor,
and $J_2=(0,0,2k/k_1)$
for the $\mathbb{Z}_{k_1}$ factor, respectively.  

Imposing the general restrictions on the $2\times 2$ matrix $X$, we are left with
a $\ints_{k_1}$ degree of freedom in the discrete torsion.  Its value can be
determined by requiring that the right-moving characters combine to $SU(2)_k$ characters, 
which means that $\mathcal{G}_k$ must be contained in the right
kernel of $X$. This yields 
\be
X = \left[ \begin{array}{ccc}
            0  &  -2/k_1        \\
            0  & -k/k_1^2   \\
           \end{array} \right] \label{matrix1}.
\ee
Inserting the simple current group $\mathcal{G}_{k,k_1}$ with choice \erf{matrix1}
for $X$ into equation
(\ref{modinv}), we get indeed the modular invariant (\ref{partition}), with $k/k_1=k_2$. Allowed choices of discrete torsion $X$ different from the one in
(\ref{matrix1}) yield other simple current constructions, which are again
consistent CFTs. One may wonder whether those theories
possess a sensible target space interpretation as well, which
would lead to a geometric interpretation of discrete torsion.

Since the $U(1)$ characters obey $\chii^{U(1)}_{n}(\tau) = \chii^{U(1)}_{-n}(\tau)$ as 
functions of the modular parameter $\tau$, the partition function of the lens space 
CFT is invariant under $n'\mapsto -n'$, that is 
$Z\left( \mathcal{L}_{k,k_1} \right) (\tau) = Z\left( \mathcal{L}_{k,k_2} \right) (\tau)$.  
In CFT terms this involutive action on the left-movers is just a $U(1)$ charge conjugation; 
geometrically, it acts in the appropriate way on the radial parameter of the lens space 
metric and on the string coupling so as to interpret it as a T-duality \cite{MMS1}.
It has already been observed in \cite{GPS} that the spectrum described by 
\erf{partition} contains winding states.
Since the size of the orbifold group that is modded out to get the lens space from
the $SU(2)$ manifold is $k_1$, 
it is tempting to interpret \cite{MMS1} the combination $(n{-}n')/2$, which
is defined mod $k_1$, as winding number and the combination $(n{+}n')/2$, defined 
mod $k_2$, as momentum. Then the transformation $n'\mapsto -n'$ amounts to interchanging
winding number and momentum, supporting its interpretation as a T-duality.
Note that -- as in the case of tori, and unlike in the case of $SU(2)$ -- the presence 
of winding states prevents us from having an interpretation
of the space of lowest weight states as (a truncation of) the space of functions
on the manifold $L_{k_1}$.

\medskip

The action by the fusion product of the simple current group $\mathcal{G}_{k,k_1}$ on the 
primary fields of the $PF_k\times U(1)_k$ theory
may have fixed points. This happens iff 
\be
K:=(k,0,0)\sim (0,k,0)= (0,2,2)^{k/2} (0,0,2k_2)^{k_1 /2}\in \mathcal{G}_{k,k_1},
\label{K}
\ee 
which requires 
$k_1$, and hence $k$, to be even.  Then 
$(0,k,0)*(k/2,n,m)$$=(k/2,n{+}k,m)\sim (k/2,n,m)$, where the last step is the field 
identification of $PF_k$. The field $(k/2,n,m)$ appears with multiplicity two in the partition function
iff both $k_1$ and $k_2$ are even.


\sect{The boundary states and annulus amplitudes}

The boundary states for CFTs of simple current type have been built in 
\cite{FHSSW}. More precisely, in \cite{FHSSW} only the case of trivial discrete torsion 
was studied, because it allows for discussing also amplitudes on non-orientable
world sheets. However, the relevant results from \cite{FHSSW} are also applicable 
for non-trivial discrete torsion \cite{huis}.

\paragraph{Boundary blocks}

We first describe the boundary blocks, or Ishibashi states. 
They are three-point conformal blocks with insertions 
of a primary field $\lambda$, its charge conjugate $\lambda^\dagger$ and some simple current $F$, 
of which $\lambda$ is a fixed point. We denote $\mathcal{G}$ the simple current group and 
$S_{\lambda}\subseteq \mathcal{G}$ is the stabilizer of $\lambda=(j,n,m)$. The boundary blocks thus
correspond to primaries that are combined  
with their charge conjugate in the partition function of the extended theory. 
Applying the results of
\cite{FHSSW,huis}, it follows that in terms of the $PF \otimes U(1)$ theory 
they are labelled by pairs $(\mu,F)$ with $F\in S_{\mu}$ that satisfy the requirement
\be
Q_{J}(\mu)+X(J,F)\in \ints,~~ \forall~ J\in \mathcal{G}.
\label{requirement}
\ee
Note that here the choice of discrete torsion enters explicitly. 

In our case, all the stabilizers are trivial, that is $S_\mu=\{\Omega\} \equiv \{(0,0,0)\}$, 
except when $k_1$ is even in which case we have 
$S_{(k/2,n,m)}=\{\Omega, K\} \cong \ints_2$, generated by
$K=(0,k,0)$ as given in (\ref{K}). 
We first concentrate on the trivial element $\Omega$ of the stabilizers. We have 
$X(J,\Omega)=0$ and the requirement (\ref{requirement}) is 
$Q_J(j,n,m)=0,~~ \forall~ J\in \mathcal{G}$. This means $m=0$ mod 
$k_1$ and $m=n$ mod $k$. Thus we have Ishibashi labels of the type
\be
(\mu,\Omega)~~~\mbox{with}~~~\mu=(j,rk_1,rk_1)~~~\mbox{and}~~~\Omega=(0,0,0),
\label{ishi1}
\ee
with $0\,{\leq\,} r\,{<\,}2k_2$ and $0\,{\leq\,} j \,{\leq\,} k$.
For a pair $(\mu,K)$ with $\mu=(k/2,n,m)$ the requirement 
(\ref{requirement}) is $n=m$ mod $k$ and $m=-k/2$ mod $k_1$. 
The corresponding Ishibashi labels are
\be
(\mu,K)~~~\mbox{with}~~~\mu=(k/2,k/2+rk_1,k/2+rk_1)~~~\mbox{and}~~~K=(0,k,0)\sim (k,0,0),
\label{ishi2}
\ee
with $0\,{\leq\,} r\,{<\,}2k_2$.
When $k_1$ is even we have $2k_2$ Ishibashi labels of this kind. 
The total number of Ishibashi labels is then
\be 
\#I=\left\{ \begin{array}{ll}
	(k+4)k_2 & \mbox{if}~~2|k_1 \\[2mm]
	(k+1)k_2 & \mbox{else} ~.
	\end{array}
	\right.
\label{ishibashi}
\ee

We label the boundary blocks by $| A; j,r,F \rangle\rangle=
| A; j,rk_1 \rangle\rangle^{PF_k}| A; rk_1\rangle\rangle^{U(1)_k}$ with 
$F=\Omega,K$ for the boundary blocks in (\ref{ishi1}) and (\ref{ishi2}) respectively. 
Following \cite{MMS1}, we use the label $A$ to indicate that they preserve the full 
$PF \otimes U(1)$ symmetry, as opposed to the ones discussed below, labelled $B$.
We normalize the A-type boundary blocks as
\be
\langle\langle A; j',r', F' | q^{L_0 \otimes {\bf 1} +{\bf 1}\otimes  L_0 - c/12}
| A; j,r, F \rangle\rangle =
     \delta_{jj'}\delta_{rr'} \delta_{FF'}\, \chi_{j,rk_1}^{PF_k}(q^2)\, \chi_{rk_1}^{U(1)_k} (q^2).
\ \ \ \ 
\ee


\paragraph{Boundary states}

The boundary states are labelled by $\mathcal{G}$-orbits on the set of 
all chiral labels of the unextended $PF \otimes U(1)$ theory, possibly with
multiplicities. More concretely, they are labelled by pairs $[\rho,\psi_\rho]$ 
with $\rho$ a representative of a $\mathcal{G}$-orbit and $\psi_{\rho}$ 
a $C_{\rho}$-character. Here $C_{\rho} \subseteq S_{\rho}$, the 
central stabilizer, is a subgroup in the stabilizer of square index \cite{FS1}. 
Since in the case at hand, the stabilizer has at most two elements, 
it follows that $C_{\rho}=S_{\rho}$ for all $\rho$. For cyclic groups, 
the values of the character are $|C_{\rho}|$-th roots of unity, 
hence they are signs for our considerations. 
Orbits whose fields do not appear in the torus partition function correspond to boundary 
conditions that break at least part of the (maximally extended)
bulk symmetry; 
they cannot be obtained
by the procedure of averaging Cardy boundary conditions over the orbifold group,
which is e.g.\ used in \cite{MMS1}.

In all cases except $2|k_2,2{\nmid}k_1$, 
the $\rho$ labels for the orbits can be represented by
\be
\rho=(j,n,n+s)~~~\mbox{with}~~~0\leq j \leq \lfloor k/2\rfloor,
\label{orbit}
\ee
where $\lfloor k/2\rfloor$ is the integer part of $k/2$. The dummy index 
$n\in\{0,1\}$ appearing here is chosen so that $2|j{+}n$, and $0\leq s \leq 2k_2{-}1$. 
We will sometimes condense the notation and supress the dummy 
index $n$, and instead label boundary states by $[j,s,\psi]$, with the 
character displayed only when it is nontrivial. All $PF \otimes U(1)$ fields lie in 
such an orbit; the only subtlety is that some of the orbits (\ref{orbit}) can actually be 
identical. This happens iff the simple current 
$(0,2,2)^{k/2}(0,0,2k_2)^{-\frac{k_1-1}{2}}=(0,k,k_2)$ is in $\mathcal G$, 
which due to field identification acts as $(0,0,k_2)$ on the label $(k/2,n,s)$. 
For this current to appear we need $2{\nmid} k_1$ and $2|k$. In that case, 
the orbits are labelled by $[j,n,n{+}s]$ with $0\leq {j} < k/2$, 
and $0\leq s \leq 2k_2{-}1$, and there is a 
second kind of orbits labelled by $[k/2,n,n{+}s]$ now with $0\leq s \leq k_2{-}1$. One can check 
that the number of boundary states equals the number of Ishibashi 
states, as predicted from the general theory (see e.g.\ \cite{FS1,bifs}).
      
For any simple current construction, the boundary coefficients that appear 
in the expression $|A,[\rho,\psi_\rho]\rangle 
= \sum_{j,r,F}B_{(\mu,F),[\rho,\psi]}|A;j,r,F\rangle\rangle$ of the boundary 
states in terms of Ishibashi blocks are given by \cite{FHSSW,huis}
\be
B_{(\mu,F),[\rho,\psi]}=\sqrt{\frac{|\mathcal{G}|}{|S_{\rho}||C_{\rho}|}}\,
\frac{\alpha_F S^{F}_{\mu,\rho}}{\sqrt{S_{\Omega,\mu}}}\,\psi(F)^*,
\label{bc}
\ee
where the indices $\mu=(jmn),\rho=(j'm'n')$ are $PF \otimes U(1)$ indices and 
$\alpha_F$ is a phase, in the present case an eighth root of unity. It can be choosen to be $\alpha_K=\eE^{\ii\pi/4}$ for $F=K$ and $4{\nmid} k$, and $\alpha_F=1$ in all other cases. $S^{F}$ 
is the modular transformation matrix for 1-point blocks with insertion $F$ 
on the torus, in particular $S^{\Omega}=S$. A current $F$ gives rise to 
nontrivial matrix elements $S^F_{\mu\rho}$ only if both $\mu$ and $\rho$ are 
fixed by $F$ \cite{FSS1}.

The $PF \otimes U(1)$ - $S$ matrix reads
\be
S_{(jmn),(j'm'n')}^{PF \otimes U(1)}
=2\,S_{j,j'}^{SU(2)}S_{n,n'}^{U(1)}(S_{m,m'}^{U(1)})^*,
\ee
with
\be
S_{n,n'}^{U(1)}=\frac{1}{\sqrt{2k}}\eE^{-\frac{\ii\pi nn'}{k}},
~~~~~~~
S_{j,j'}^{SU(2)}=\sqrt{\frac{2}{k+2}}\,\sin \left(
\frac{(j'+1)(j+1)}{k+2}\pi
\right).
\ee
In case $k_1$ is odd, the stabilizers are all trivial and $\psi(F)=\psi(\Omega)=1$. 
All boundary states $|A,j',s'\rangle$ that are related to trivial characters 
can be expressed as
\begin{eqnarray}
|A,j',s'\rangle
=\sqrt{k_1}\sum_{j=0,1,...,k \atop{r=0,1,...,2k_2-1\atop{2|(j+rk_1)}}}
\frac{S_{j,j'}^{SU(2)}}{\sqrt{S_{0,j}^{SU(2)}}}\,
\eE^{-\frac{\ii\pi rk_1s'}{k}}
|A;j,r,\Omega \rangle\rangle.
\label{boundary}
\end{eqnarray}
In case $k_1$ is even, there are nontrivial stabilizers and we must take 
into account the character $\psi_{\rho}$ of $C_{\rho}$. The boundary coefficients (\ref{bc})
depend on whether the stabilizing current $J$ is trivial or not. For $J=\Omega$, 
the boundary coefficients are just as above,
so for $0\leq j' < \lfloor k/2\rfloor$, 
we can express the boundary state in terms of boundary blocks again as 
in equation (\ref{boundary}). So in case $j'=k/2$ we have also a summation 
over the Ishibashi states appearing in (\ref{ishi2}),
\be
\label{frac}
|A,k/2,s,\psi\rangle&=&\frac{\sqrt{kk_1}}{2}\Bigg(
\sum_{j=0,2,...,k \atop{r=0,1,...,2k_2-1}}
\frac{S_{(j,rk_1,rk_1),(k/2,n,n+s)}}{\sqrt{S_{\Omega,(j,rk_1,rk_1)}}}\,
|A;j,r,\Omega \rangle\rangle
\\
&&+\sum_{r=0,1,...,2k_2-1}\!
\frac{\alpha_K\psi(K)S^{K}_{(k/2,k/2+rk_1,k/2+rk_1),(k/2,n,n+s)}}
{\sqrt{S_{\Omega,(k/2,k/2+rk_1,k/2+rk_1)}}}\,
|A,k/2,r,K \rangle\rangle
\Bigg),
\nonumber
\ee
where we omitted the superscript $PF \otimes U(1)$ on the S-matrices because this
can be recognized from the form of the (multi-)labels that appear of the indices. The $S^{K}$ matrix appearing in \erf{frac} 
be factorized in its $SU(2)$ and $U(1)$ parts just like the ordinary $S$ matrix,
\be
S^{K}_{(k/2,n',m'),(k/2,n,m)}=\frac{1}{k}\, S^{K,SU(2)}_{k/2,k/2}\,\eE^{\frac{\ii\pi}{k}(n'n-m'm)}
=\frac{1}{k}\, D\,\eE^{\frac{\ii\pi}{k}(n'n-m'm)}
\ee
with \cite{FSS2} $D=\eE^{-3\pi\ii k/8}$. 
These branes are called ``fractional" branes, reflecting the additional factor of 
$\frac12$ in (\ref{frac}) (which then also arises in the annulus amplitude 
(\ref{fracamp})), as opposed to the ones which involve only a 
summation over the boundary blocks in (\ref{ishi1}).

\paragraph{B-type branes}

{}Recall that from inspection of the lens space partition function (\ref{partition}) and the 
relation $\chii^{U(1)}_{n}=\chii^{U(1)}_{-n}$, one sees that the T-dual (along the 
Cartan subalgebra $U(1)$ in $SU(2)$) of the $\mathcal{L}_{k,k_2}$ theory is the
$\mathcal{L}_{k,k_1}$ theory.  The T-duals of the A-type branes constructed
above in $\mathcal{L}_{k,k_2}$ give a new type of branes in $\mathcal{L}_{k,k_1}$, 
called B-type branes \cite{MMS1}.
These can be studied by
regarding the boundary blocks of $\mathcal{L}_{k,k_1}$ as tensor products 
of the boundary blocks of the parafermion and free boson theories.
Indeed, since T-duality amounts to changing the sign of the 
$U(1)$ label $n$ in the left-moving field labelled $(j,m,n)$, 
leaving the $PF$ labels unchanged, the B-type boundary blocks 
can be written as
\be
|B;j,rk_2,rk_2, \Omega \rangle\rangle ^{\mathcal{L}_{k,k_1}}
=|A;j,rk_2\rangle\rangle^{PF_k}|B;rk_2\rangle\rangle^{U(1)_k}.
\ee
where the label $B$ on the right is to remind us that we have to
switch sign on the left-moving momentum,
$|B;rk_2 \rangle\rangle^{U(1)_k} := |{-}rk_2,rk_2\rangle\rangle^{U(1)_k}$. These will be seen to give nonvanishing contributions only for $r=0$ mod $k_1$ in section \ref{geom}.
The B-type branes in the $\mathcal{L}_{k,k_1}$ theory are T-dual to A-type branes in 
$\mathcal{L}_{k,k_2}$. This gives for the nonfractional branes
\begin{eqnarray}
|B;j',s'\rangle^{\mathcal{L}_{k,k_1}}
=\sqrt{k_2}\sum_{j=0,1,...,k \atop{r=0,1,...,2k_1-1\atop{2|(j+rk_2)}}}
\frac{S_{j,j'}^{SU(2)}}{\sqrt{S_{0,j}^{SU(2)}}}\,
\eE^{-\frac{\ii\pi rk_2s'}{k}}
|A;j,rk_2\rangle\rangle^{PF_k}|B;rk_2\rangle\rangle^{U(1)_k},
\label{boundary}
\end{eqnarray}
with the range of $s'$ as before but $k_1$ interchanged with $k_2$. 
Likewise, for the fractional branes which arise iff $2|k_2$, we get from performing T-duality on 
(\ref{frac}) expressed with $k_1$ interchanged with $k_2$,
\be
&& \!\!\!\!\!\!\!\!\!\!\!\!\!\! |B;k/2,s,\psi\rangle^{\mathcal{L}_{k,k_1}}=\frac{\sqrt{kk_2}}{2}\Bigg(
\sum_{j=0,2,...,k \atop{r=0,1,...,2k_1-1}}
\frac{S_{(j,rk_2,rk_2),(k/2,n,n+s)}}{\sqrt{S_{\Omega,(j,rk_2,rk_2)}}}
|A;j,rk_2\rangle\rangle^{PF_k}|B;rk_2\rangle\rangle^{U(1)_k} 
\\
&&\quad +\!\! \sum_{r=0,1,...,2k_1-1}\!\!
\frac{\alpha_K\psi(K)S^{K}_{(k/2,k/2+rk_2,k/2+rk_2),(k/2,n,n+s)}}
{\sqrt{S_{\Omega,(k/2,k/2+rk_2,k/2+rk_2)}}}
|A;\mbox{$\frac k2$}{+}rk_2\rangle\rangle^{PF_k}|B;\mbox{$\frac k2$}{+}rk_2\rangle\rangle^{U(1)_k}
\Bigg).
\nonumber
\ee

\paragraph{SU(2)}

Now let $k_1=1$, $k_2=k$ in which case we recover the $SU(2)_k$ theory. 
We see in (\ref{frac}) and in (\ref{boundary}) that we 
get more boundary blocks and boundary states than discussed in \cite{MMS1}, who found 
the boundary states with $s=0$, which gives the Cardy formula 
for $SU(2)$ symmetry preserving boundary states. Note that imposing $s=0$
amounts to a restriction to a subspace in the space of boundary blocks. 
In contrast, when allowing for all boundary blocks obtained above, we get 
all $PF \otimes U(1)$ preserving boundary states; except for $s=0$, they
do not preserve the full $SU(2)$ symmetry.

Now the B-branes in $SU(2)=\mathcal{L}_{k,1}$ are T-dual to the A-branes of 
$\mathcal{L}_{k,k}$. For the case $2{\nmid} k$, the results of \cite{MMS1} 
for B-branes are recovered by setting $\eta:=(-1)^{(s'+j')}$. The 
most interesting $B$-branes are the fractional ones, which are not discussed 
in detail in \cite{MMS1}. They correspond to the fixed point arising when $2|k$, 
\begin{eqnarray}
|B;k/2;s=0,1;\psi\rangle_C&=&\frac{k}{2}\Bigg(
\sum_{j=0,2,...,k \atop{r=0,1}}
\frac{S_{(j,rk,rk),(k/2,n,n+s)}}{\sqrt{S_{\Omega,(j,rk,rk)}}}
|j,r,\Omega\rangle\rangle
\\
&+&\sum_{r=0,1}
\frac{\alpha\psi S^{K}_{(k/2,k/2+rk,k/2+rk),(k/2,n,n+s)}}
{\sqrt{S_{\Omega,(k/2,k/2+rk,k/2+rk)}}}
|k/2,r,K\rangle\rangle \Bigg).
\nonumber \end{eqnarray} 
For $k_1 = 2$, i.e.\ $SO(3)_k$, our results agree with those of 
\cite{FFFS,couch}.


\paragraph{Remark}
We emphasize the fact that the simple current construction yields directly the correct boundary
states, in particular the appropriate set of boundary blocks and reflection coefficients.
In other approaches, like the one of \cite{MMS1}, the reflection coefficients are 
determined by imposing the NIMrep properties of the annulus coefficients. 
This is a necessary, but not sufficient condition for the CFT to make sense;
indeed, many NIMreps are known which do not appear in any consistent CFT
(see e.g.\ \cite{gann17}).
The simple current construction can be shown to yields NIMreps that are physical,
i.e.\ do belong to a consistent CFT \cite{FRS}; thus the boundary conditions studied
here (and hence in particular those also discussed in \cite{MMS1}) are indeed physical.

As already mentioned in the introduction, starting from $PF \otimes U(1)/\ints_2$ instead 
of $PF \otimes U(1)$, with $U(1)/\ints_2$ the $\ints_2$ orbifold of the free boson,
would allow one to describe the $B$-type boundary conditions on the same footing
as the A-type conditions. Unfortunately, while both the extension from 
$PF \otimes U(1)/\ints_2$ to $PF \otimes U(1)$ and the construction of $\mathcal{L}_{k,k_1}$
from $PF \otimes U(1)$ are simple current construction, this is no longer true for the
construction of $\mathcal{L}_{k,k_1}$ directly from $PF \otimes U(1)/\ints_2$, because
in terms of the latter theory, the simple currents in $\mathcal{G}_{k}$ correspond to
fields of quantum dimension 2. Describing A- and B-type conditions in this manner
will therefore require to apply the results of \cite{FRS} on such more general constructions.


\paragraph{Annulus coefficients}
In the closed string channel the annulus amplitude with boundary conditions $a,b$ is
\be
A_{a}^{~b}(t)=\langle a|\eE^{-\frac{2\pi}{t}(L_0 \otimes {\bf 1} +{\bf 1}\otimes L_0 - c/12)}|b\rangle.
\ee
In the open string channel, we can expand the amplitude in terms of characters as
$A_{a}^{~b}(t)=\sum_{\nu}A_{a}^{\nu~b} \chii_{\nu} (\eE^{\frac{it}{2}})$.
For a simple current construction, the annulus coefficients $A_{a}^{\nu~b}$ appearing in this expansion are given by \cite{huis}
\be
A_{[a,\psi_a]}^{\nu~[b,\psi_b]}
=\sum_{(\mu,J)}B_{(\mu,J)[a,\psi_a]}B^*_{(\mu,J^c)[b,\psi_b]}S^{\nu}_{\mu}.
\ee
This formula depends on the choice of discrete torsion through the restrictions 
on the summation over Ishibashi labels.
In case $k_1$ is odd, the result is
\be
A_{[j_a,s_a]}^{(j',n',m')~[j_b,s_b]}=\delta^{2k_2}(s_a{-}s_b{+}n'{-}m')\, N_{j_aj_b}^{j'},
\ee
where $N_{j_aj_b}^{j'}$ are the $SU(2)_k$ fusion rules.

Now we consider the case $k_1$ is even. If none of the orbits $a,b$ is a fixed point, we get
\be
A_{[j_a,s_a]}^{(j',n',m')~[j_b,s_b]}&=&\delta^{2k_2}(s_a{-}s_b{+}n'{-}m')
\left(N_{j_aj_b}^{j'}+N_{j_aj_b}^{k/2-j'}\right).
\ee
In case precisely one of the boundary labels is a fixed point, the computation is similar, since
the appearing $S^K$ matrix elements vanish, and
$A_{[k/2,s_a,\psi_a]}^{(j',n',m')~[j_b,s_b]}=\delta^{2k_2}(s_a{-}s_b{+}n'{-}m')N_{k/2,j_b}^{j'}$
(which does not depend on $\psi_a$). In case both orbits $a,b$ are fixed points we have to 
include both types of Ishibashi labels in the summation. We get
\be
A_{[\mbox{$\frac k2$},s_a,\psi_a]}^{(j',n',m')\,[\mbox{$\frac k2$},s_b,\psi_b]}
=\frac{1}{2}\,\delta^{2k_2} (s_a{-}s_b{+}n'{-}m')\!\left(N_{\mbox{$\frac k2$},
\mbox{$\frac k2$}}^{j'}
{+} \ii^{s_a{-}s_b+n'-m'}\psi_a\psi_b
\sin((j'{+}\mbox{$\frac 12$})\pi)\right).
\label{fracamp}
\ee
Note that this is always a non-negative integer, since the fusion rule $N_{k/2,k/2}^{j'}$ 
is nonvanishing exactly when $\sin(\pi(j'{+}1)/2)$ is, 
and $2|(s_a{-}s_b{+}n'{-}m')$ due to the Kronecker delta in the prefactor. 

To sum up, the annulus coefficients are
\[ A_{a}^{\nu~b}=\left\{ \begin{array}{ll}
	\delta\times N_{j_aj_b}^{j'} & \mbox{if}~~2{\nmid} k_1 \\[4pt]
	\delta\times\left(N_{j_aj_b}^{j'}+N_{j_aj_b}^{k/2-j'}\right)
	& \mbox{if}~~2| k_1 ~~\mbox{and}~~ j_a,j_b\neq k/2 \\[4pt]
	\delta\times N_{k/2,j_b}^{j'}
	& \mbox{if}~~2| k_1 ~~\mbox{and}~~ j_b\neq k/2, ~~j_a= k/2\\[4pt]
	\delta\times\frac{1}{2}\,\delta^{1}(j')\left(1
	+ \psi_a\psi_b
	(-1)^{j'+\frac{1}{2}(s_a{-}s_b+n'-m')}\right)
	& \mbox{if}~~2| k_1 ~~\mbox{and}~~ j_a,j_b= k/2
	\end{array}
	\right.
\]
with $\delta \equiv \delta^{2k_2}(s_a{-}s_b{+}n'{-}m')$, $\nu \equiv (j',n',m')$, 
$a \equiv [j_a,s_a,\psi_a]$ and $b \equiv [j_b,s_b,\psi_b]$. 
This generalizes the results of \cite{MMS1} who considered 
boundary states with $s=0$, and where the details on the last amplitude where 
not carried out explicitly. 
All amplitudes 
depend only on the difference $s_a{-}s_b$, which is consistent with the results 
 below that indicate that the branes with $s\neq 0$ correspond to averaging over 
the $\ints_{k_1}$ orbits of twisted \cite{FFFS} conjugacy classes in the covering space.


\sect{On the geometry of the branes} \label{geom}

In this section, we discuss the geometry of the branes following \cite{FFFS}.
A brane on $SU(2)$ is a linear combination of Ishibashi states, or boundary blocks $B_{j}$. These are linear functionals $B_j{:}~H_j\otimes H_{j^\dagger}\to \complex$ and we can restrict this action to the horizontal submodules $\bar H_j\otimes \bar H_{j^\dagger}\subset H_j\otimes H_{j^\dagger}$. In the large $k$ limit, all $j$ are allowed and we can  identify the boundary blocks 
with functions $\tilde B_{j}$ on the group manifold through the Peter-Weyl isomorphism 
\be
\tilde B_{j'}(g)=\sum_{j,m_L,m_R} \sqrt{\tfrac{j+1}{V}} B_{j'}(v_{m_L}^j\otimes \tilde v_{m_R}^j)
\,\langle v_{m_L}^j | R^j(g) |\tilde v_{m_R}^j\rangle ,
\ee
with $\{v_{m}^j\}$ a basis of the $\su(2)$ module $\bar H_j$ with highest weight $j$, and $V$ the volume of the group manifold $SU(2)$. Recall that the lens space $L_{k_1}$ is the set of equivalence classes of $SU(2)$ group elements with the equivalence relation $g\sim \eE^{\frac{2\pi\ii}{k_1} H}g$.
A function on $SU(2)$ is independent of the choice of representative of such a class, and therefore a function on the lens space, iff $f (v_{m_L}^j\otimes \tilde v_{m_R}^j)\neq 0$ only for $m_L=0$ mod $k_1$. For convenience we then still write a group element, instead of an equivalence class, as argument for such a function. In geometric terms, the isomorphism will give us the profile of the branes as probed by the tachyons on the target space. Similar expressions result
when the graviton, dilaton and Kalb-Ramond fields are used as probes; qualitatively,
they all give the same profile \cite{FFFS}.  

The Ishibashi states for the lens space CFT are expressed as $B_{jmn}=B_{jm}^{PF}B_{n}^{U(1)}$ in terms of $PF$ and $U(1)$ Ishibashi states. We decompose the functions on $SU(2)$, and the corresponding states, into functions (states) on $SU(2)/U(1)$ and $U(1)$,  $v^j_n\otimes \tilde v^j_m=w^j_{(m+n)/2}\,e_{n,-m}$, such that $B^{U(1)}_q(e_{n,-m})=\delta_{q,n}\delta_{q,m}$, and 
\be
B_{j',rk_1}^{PF}(w^j_{(m_L+m_R)/2})=\delta_{j,j'}\delta^{4k} _{2rk_1,m_L+m_R}+\delta_{j,k-j'}\delta^{4k} _{2rk_1-2k,m_L+m_R},
\label{AI}
\ee
where field identification of the boundary blocks is taken into account. (In the special case $j=k$, we have $k+1$ states in the $SU(2)$ representation whereas the range of Ishibashi only allows $k$ states. Accordingly one linear combination of $v^k_{k}\otimes \tilde v^k_{-k}$ and $v^k_{-k}\otimes \tilde v^k_{k}$ is annihilated by all Ishibashi states.)
The shape of the regular A-branes is
\be
\tilde \mathcal B_{j,s}(g)=\sqrt{k_1}\!\!\!\!\!\!\!\!\!\!
\sum_{j'=0,1,...,k \atop{r=-k_2,-k_2+1,...k_2-1\atop{2|(j'+rk_1)}}}
\!\!\!\!\!\!\!\!\!\!
\hat S_{j,j'}
D^{j'}_{rk_1,rk_1}(g_s)
=\tilde \mathcal B_{j,0}(g_s)~,
\ee
where we introduce the shorthands
\be
~~~\hat S_{j,j'}:= S_{j,j'}^{SU(2)} \sqrt\frac{j'+1}{S_{0,j'}^{SU(2)}V}
~~,~~~~~
g_s:=\eE^{-\frac{\ii\pi s}{k}H}g \label{gS}
~,~~~\mbox{and}~~~~~
D^{j}_{m,n}(g):=\langle v_{m}^j | R^j(g) |\tilde v_{n}^j\rangle~.
\ee
The latter matrix element vanishes unless $|m|,|n|\leq j$. We interpret $s$ as parametrizing a rotation of the brane. Since $0\,{\leq}\, s \,{<}\, 2k_2$ except for the exceptional case, we see that the exponent in $\eE^{\frac{\ii\pi s}{k}H}$ has the range $0\leq \frac{\ii\pi s}{k}H < \frac{2\ii\pi }{k_1}H$, where the rightmost expression is the one we are orbifolding. 

For the fractional branes, we also need the fractional Ishibashi states, where the analogue to the second term in \erf{AI} gives a nonzero contribution. The shape of the fractional brane is
\be
\label{frac}
\tilde \mathcal B_{s,\psi}(g)=\frac{\sqrt{kk_1}}{2}\!\!\!\!\!\!\!\!\!\!\!\!
\sum_{j'=0,2,...,k \atop{r=-k_2,-k_2+1,...,k_2-1}}\!\!\!\!\!\!\!\!\!\!\!\!
\hat S_{j,j'}
D^{j'}_{rk_1,rk_1}(g_s)
+ \sqrt{\tfrac{k_1}{V}}\, \eE^{-3\pi\ii \frac k8}\mbox{$\left( \frac{k+2}{2}\right)^{\frac 14}$}
\,\psi\, \delta^{2}_{k_2,0}\!\!\!\!\sum_{r=\pm k/2}
D^{k/2}_{r,r}(g_s),
\ee
with $g_s$ and $\hat S_{j,j'}$ as in \erf{gS}.
The two fractional branes have the same shape unless $2|k_2$, due to $\delta^{2}_{k_2,0}$. 

For the A- branes on $SU(2)$, i. e. the lens space with $k_1{=}\,1$, the shape reduces to
$\tilde \mathcal B_{j,s}(g)=\sum_{j'}
\hat S_{j,j'}\,
\chii_{j'}(g_s)$. For $s=0$, these branes are the standard Cardy branes described in \cite{MMS1}\footnote{With one term missing in the $j'=k$ character due to problems with matching the space of functions on the lens space to the set of Ishibashi labels.}. At finite level, the support of the profile, which one would like to interpret as
the world volume of the brane, is in fact the whole target space. But it is peaked around
a conjugacy class (if $s=0$) respectively a tilted conjugacy classes (if $s\ne 0$) of $SU(2)$,
so that at finite level one can think of it as a smeared brane at the (tilted) conjugacy class
\cite{Aleks-Schom,FFFS,staN5,Gawe}. 

For the B-branes in lens spaces, we have to use the B-type $U(1)$ Ishibashi states, acting as
$B^B_{rk_2}(e_{n,-m})=\delta_{rk_2,n}\delta_{-rk_2,m}$. 
The shape of the nonfractional B-branes is given by
\be
\tilde \mathcal B^B_{j,s}(g)=\sqrt{k_2}
\!\!\!\!\!\!\!\!\sum_{j'=0,1,...,k \atop{r=0,k_1~~2|(j'+rk_2)}}\!\!\!\!\!\!\!\!
\hat S_{k/2,j'}\,
D^{j'}_{rk_2,rk_2}(g_s)=\tilde \mathcal B^B_{j,0}(g_s),
 \ee
 with $s=0,...,2k_1{-}1$, or $s=0,...,k_1{-}1$ in the case $2|k_1,2{\nmid}\, k_2$. In particular, for $SU(2)$ it is
 \be
 \tilde \mathcal B^B_{j,s}(g)=\sqrt{k}
 \!\!\sum_{j'=0,2,...,2[k/2]}\!\!
 \hat S_{j,j'}\,
 D^{j'}_{0,0}(g_s)
 +\sqrt{k}
 \hat S_{j,k}\,
 D^{k}_{k,k}(g_s)
 \ee
 The fractional B-brane that arises when $2|k_2$ has the shape
\be
\label{Bfrac}
\tilde \mathcal B ^B _{s,\psi}(g)
=\frac {\sqrt{k_2}}2
\!\!\!\!\sum_{j'=0,2,...,k \atop{r=0,k_1}}\!\!\!\!
\hat S_{k/2,j'}
D^{j'}_{rk_2,rk_2}(g_s)
+\delta^2_{k_1,0}
\,\psi\,  \eE^{\frac{-3\pi\ii}8} 
D^{k/2}_{0,0}(g_s)\sqrt\frac{k_2}{V}\left(\tfrac {k+2}{2}\right)^{\frac 14}~.
\ee
By power-counting in $k$, we see the $\psi$ dependent terms in \erf{frac}
 and \erf{Bfrac} do not contribute to the shape of the fractional branes in the 
 limit of large level.

\paragraph{Flux stabilization of the branes}

The flux quantization mechanism for A-branes in $SU(2)$ \cite{BDS} can 
be used to establish similar results for A-branes in lens spaces. A nonfractional A-brane in the lens space 
$L_{k_1}= SU(2)/\mathbb{Z}_{k_1}$ is a projection to $L_{k_1}$ 
of a union 
of (twined) conjugacy classes in $SU(2)$. In the large level limit, the shape of the fractional branes can also be seen as such projections and we can do a similar discussion as the one below. 
The $SU(2)$ flux stabilization mechanism applies independently to each of
the pre-images of the lens space brane.
The stabilizing fields in general depend on the pre-image
just by a multiplicative factor.

The variational problem in the lens space can now be rephrased as a variational problem
in $SU(2)$.  The Born-Infeld action for the brane in the lens space is proportional to
the sum of the Born-Infeld terms describing the branes in the covering space.
Each of the image $SU(2)$ branes is a stable solution to the Born-Infeld
variational problem. 
So, in particular, the value of the action is stable under fluctuations that 
survive the projection to the lens space.
Hence $|A;j,s\rangle_{k_1}$ constitutes a solution of the Born-Infeld equations of 
motion for the lens space. In the particular example of $SO(3)=SU(2)/\ints_2$, these results are illustrated 
in \cite{couch}.


\vskip3em

{\bf Acknowledgements.}
We thank J\"urgen Fuchs and Christoph Schweigert for introducing us to the subject and for many 
helpful comments. P.B. is supported by the grant SFRH/BD/799/2000 of FCT (Portugal).


\end{document}